\documentclass[11pt]{article}
\usepackage{amsmath,amsfonts,amsthm,amssymb}

\newcommand\version{April 17, 2006}
\numberwithin{equation}{section}

\newcommand\Z{{\mathbb Z}}

\newcommand\C{{\mathbb C}}

\newcommand\R{{\mathbb R}}

\newcommand\half{\mbox{$\frac 12$}}
\newcommand\rhov\varrho

\newcommand\const{{\rm const.\, }}
\newcommand\Tr{{\rm Tr}}
\newcommand\tr{{\rm tr}}

\newcommand\Hh{{\cal H}}

\newcommand\per{{\rm per}}
\newcommand\vu{\nu}

\renewcommand\kappa\varkappa
\renewcommand\rho\varrho

\newtheorem{thm}{THEOREM}

\newtheorem{lem}{Lemma}

\begin{document}

\markboth{\scriptsize{\version}}{\scriptsize{\version}}
\title{\bf{A Correlation Estimate for Quantum Many-Body Systems at Positive Temperature}}
\author{\vspace{5pt} Robert Seiringer
\\ \vspace{-2pt}\small{ Department of Physics, Jadwin Hall, Princeton
University,  }\\ \vspace{-2pt}\small P.O. Box 708, Princeton NJ
08544, USA. 
\\ {\small Email: \texttt  {rseiring@math.princeton.edu}} }

\date{\small \version}
\maketitle

\begin{abstract}
  We present an inequality that gives a lower bound on the expectation
  value of certain two-body interaction potentials in a general state
  on Fock space in terms of the corresponding expectation value for
  thermal equilibrium states of non-interacting systems and the
  difference in the free energy. This bound can be viewed as a
  rigorous version of first order perturbation theory for many-body
  systems at positive temperature. As an application, we give a proof
  of the first two terms in a high density (and high temperature)
  expansion of the free energy of jellium with Coulomb interactions,
  both in the fermionic and bosonic case. For bosons, our method works
  above the transition temperature (for the non-interacting gas) for
  Bose-Einstein condensation.
\end{abstract}

\renewcommand{\thefootnote}{${\,}$}
\footnotetext{Work partially supported by U.S. National Science
Foundation grant PHY-0353181 and by an Alfred P. Sloan Fellowship.}
\renewcommand{\thefootnote}{${\, }$}
\footnotetext{\copyright\,2006 by the author.
This paper may be reproduced, in its entirety, for non-commercial
purposes.}

\section{Introduction}

Correlations play a crucial role in quantum-mechanical many-body
systems. They result from interactions among the particles, and it is
typically very difficult to obtain information about them in a
mathematically rigorous fashion.  Approximate theories are often
arrived at by neglecting correlations, for instance in Hartree-Fock
theory for fermions. For the problem of estimating the validity of
such approximations, it is necessary to estimate the magnitude of
correlations present in the state of the interacting system.

In \cite{grafsol}, Graf and Solovej present a correlation estimate
which is applicable for the study of this problem at zero temperature,
i.e., for systems in their ground states. The inequality presented
there is motivated by earlier correlation estimates by Bach
\cite{bach} and Bach {\it et al.} \cite{bach2}.  Roughly speaking, it estimates the difference of the
interaction energy in a general state and the ground state of a
non-interacting system in terms of the difference of their
one-particle density matrices. Moreover, at least in the case of
fermions, the one-particle density matrix can be easily controlled in
terms of the total kinetic energy.  For bosonic systems, the situation
is more complicated, and the correlation estimate in \cite{grafsol} is
only applicable provided one can prove the existence of {\it Bose-Einstein
  condensation} --- in general a very difficult task for interacting
systems.

\iffalse
The result in \cite{grafsol} implies, in particular, the
following: If the one-particle density matrix of a general state is close to
the one for the ground state of a non-interacting system,
then the corresponding state is sufficiently close to the
non-interacting ground state such that the difference in interaction
energy between these to states is small (and can be rigorously
estimated). The one-particle density matrix, on the other hand, can be
easily estimated just by knowledge of the kinetic energy, at least in
the case of fermions. For bosons, the situation is more complicated,
and the correlation estimate in \cite{grafsol} is only applicable if
one can prove the existence of {\it Bose-Einstein condensation}, which
is in general impossible of interacting systems.
\fi

With the aid of the correlation estimate just mentioned, Graf and
Solovej were able to derive the first two terms in a high density
expansion of the ground state energy of fermionic jellium
\cite[Thm.~2]{grafsol} with Coulomb interactions. High density
corresponds to small coupling, and hence the result can be viewed as
rigorous estimate of the validity of first-order perturbation theory
for this system.

In this paper, we present a method that is applicable to the
aforementioned problem for systems at positive temperature. Unlike the
situation for the ground state, the knowledge of the one-particle
density matrix alone does not yield much information about
correlations present in the state. As an additional input one needs to
know that the {\it entropy} of the state is close to the maximal value
possible for given one-particle density matrix; this maximum is attained by
the corresponding {\it quasi-free state}. More precisely, we will
estimate the difference of the interaction energy of a general state
and the thermal equilibrium state of a non-interacting system in terms
of the {\it relative entropy} of these two states. This relative entropy
is related to the difference in free energy. Our
result applies to fermions at any temperature, and to bosons above the
critical temperature (for the non-interacting gas) for Bose-Einstein
condensation.

Our main correlation estimate is stated in Theorem~\ref{P1} in
Section~\ref{corrsect}.  Before describing it in detail, we present an
application of the inequality to (fermionic or bosonic) jellium with Coulomb interactions at
positive temperature. We will derive the first two terms in a high
density (and high temperature) expansion of the free energy. In the
fermionic case, this result can be viewed as the positive temperature
analogue of Theorem~2 in \cite{grafsol}.

Our estimate is general enough to be applicable to a wide range of
possible interparticle interactions. The two-body potential is
required to be positive definite and, in particular, to be
decomposable into characteristic functions of balls. In the case of
the Coulomb potential, such a decomposition was first used in
\cite{feff}. The study in \cite{hsdec} provides a criterion for the
possibility of such a decomposition for general radial functions, and
thus provides many examples of interaction potentials which our method
applies to.

\bigskip
{\it Acknowledgments.} It is a pleasure to thank Elliott Lieb and Jan Philip Solovej for stimulating and fruitful discussions.

\section{Jellium}

Jellium is a model of a charged gas of either fermions or bosons,
moving in a uniformly charged background. We assume that the whole
system is neutral (in a sense to be made precise below) and contained
in a (three-dimensional) cubic box of side length $L$, which we denote by $\Lambda$. We
work in the grand-canonical ensemble, i.e., in the (anti-)symmetric
Fock space over the one-particle space $\Hh=L^2(\Lambda;\C^n)$. Here, $n\geq 1$ denotes the
number of internal degrees of freedom, corresponding to particles of
spin $(n-1)/2$.

We denote by $\Delta$ the Laplacian on $\Lambda$ with Dirichlet
boundary conditions. We choose units such that $\hbar=1$ and $2m=1$,
with $m$ denoting the particle mass. For $\rho>0$ the background
density and $\alpha>0$ the square of the particle charge, the
Hamiltonian on Fock space is
\begin{equation}\label{ham}
H = H_0 + \alpha W\,,
\end{equation}
where, in each $N$-particle sector,
\begin{equation}
H_0 = - \sum_{i=1}^N \Delta_i
\end{equation}
and
\begin{equation}
W=- \sum_{i=1}^N \rho \int_\Lambda dy\,
  \frac{1}{|x_i-y|}  +\sum_{i<j} \frac 1{|x_i-x_j|}
+ \half
\rho^2 \int_{\Lambda\times\Lambda}dy_1\, dy_2\, \frac 1{|y_1-y_2|}\,.
\end{equation}
The last constant corresponds to the electrostatic energy of the
background charge and is added to ensure the existence of a proper
thermodynamic limit.

The quantity of interest is the free energy per unit volume at temperature
$T=\beta^{-1}$, given by
\begin{equation}\label{deff}
f^{\rm F,B}(\beta,\rho,\alpha)= - \lim_{L\to\infty} \frac 1{\beta |\Lambda|} \ln\, \Tr\,
\exp[-\beta H]\,.
\end{equation}
Here, $\Tr$ denotes the trace either over the fermionic (F) or bosonic (B) Fock space. 
Existence of the thermodynamic
limit in (\ref{deff}) was shown by Lieb and Narnhofer in \cite{liebnarn}. 
There it was also
shown that one would obtain the same result in the canonical ensemble
with charge neutrality, i.e., fixing $N$ to be $\rho |\Lambda|$.
In particular, in our grand-canonical setting it is not necessary to enforce the charge neutrality
$N=\rho |\Lambda|$ explicitly, it will be automatically satisfied (for the
average particle number). 

There are three length scales in this problem; the mean particle
distance $\rho^{-1/3}$, the thermal wavelength $\beta^{1/2}$, and the
inverse coupling constant $\alpha^{-1}$. Hence, by simple scaling,
\begin{equation}\label{scal}
f^{\rm F,B}(\beta,\rho,\alpha)=\rho^{5/3}
f^{\rm F,B}(\beta\rho^{2/3},1,\alpha\rho^{-1/3})\,.
\end{equation}
We are interested in the high density (and high temperature)
asymptotics; more precisely, in large $\rho$ for fixed
$\beta\rho^{2/3}$ (and fixed $\alpha$). By the scaling property
(\ref{scal}), this corresponds to a limit of small coupling.

For the statement of our main results, we will distinguish between the
fermionic and bosonic cases.

\subsection{Fermions}
Let $f^{\rm F}_0(\beta,\rho)$ denote the free energy (per unit volume) of a
non-interacting gas of spin $(n-1)/2$ fermions, at inverse temperature $\beta$ and average density $\rho$. It is given by
\begin{equation}\label{fnode}
f^{\rm F}_0(\beta,\rho) = \sup_{\mu\in\R} \left\{ \mu \rho- \frac n{(2\pi)^3\beta} \int_{\R^3}
dp\, \ln\left( 1 + e^{-\beta (p^2- \mu)}\right) \right\} \,.
\end{equation}
The supremum in (\ref{fnode})
is attained uniquely at some $\mu=\mu^{\rm F}_0(\beta,\rho)$. We denote the fugacity by
$z=e^{\beta\mu}$ for this value of $\mu$. Note that $z$ depends only on $\beta\rho^{2/3}$. Let
\begin{equation}\label{defg0}
\gamma^{\rm F}_0(p) = \frac 1 {z^{-1} e^{\beta p^2}+1}\,,
\end{equation}
and let $\widetilde\gamma^{\rm F}_0(x)=(2\pi)^{-3} \int dp\, \gamma^{\rm F}_0(p) e^{ipx}$ denote
its inverse Fourier transform. Note that $n\widetilde\gamma^{\rm
  F}_0(0)=\rho$.

\begin{thm}[High Density Asymptotics for Fermions]\label{T1}
As $\rho\to\infty$ and $\beta\to 0$, 
\begin{equation}\label{thmeq}
f^{\rm F}(\beta,\rho,\alpha) = f^{\rm F}_0(\beta,\rho) - \frac  {\alpha n}2 \int_{\R^3}dx\,
\frac{|\widetilde\gamma^{\rm F}_0(x)|^2}{|x|} - o(\rho^{4/3})\,,
\end{equation}
with $0\leq o(\rho^{4/3}) \leq C(\beta\rho^{2/3}) \alpha \rho^{4/3}
(\alpha\rho^{-1/3})^{1/48}$. Moreover, the function $C(\beta\rho^{2/3})$ is uniformly bounded on
compact intervals in $(0,\infty)$. 
\end{thm}

Note that, for fixed $\beta\rho^{2/3}$ (and fixed $\alpha$), the first term on the right side of (\ref{thmeq}) is
$O(\rho^{5/3})$, whereas the second term is
$O(\rho^{4/3})$. Theorem~\ref{T1} is the positive temperature analogue
of Theorem~2 in \cite{grafsol}.

We remark that (\ref{thmeq}) actually holds uniformly in
$\beta\rho^{2/3}$ for bounded $1/(\beta\rho^{2/3})$, with possibly a
worse exponent in the error term than the one given in
Theorem~\ref{T1}. I.e., it is uniform as the ground state is
approached. This can be proved by supplementing our lower bound with a
bound obtained with the method in \cite{grafsol} at very low
temperatures. We do not give the details here, but refer the reader to
\cite{fermiT} where a similar argument was given in the case of a
dilute Fermi gas with short-range interactions.
 
\subsection{Bosons}
For bosons we have to restrict our attention to temperatures bigger
than the critical temperature (for the non-interacting gas) or,
equivalently, to $\rho< \rho_c(\beta)\equiv n
(4\pi\beta)^{-3/2}\sum_{\ell\geq 1}\ell^{-3/2}$.  Let $f^{\rm 
  B}_0(\beta,\rho)$ denote the free energy (per unit volume) of a
non-interacting gas of spin $(n-1)/2$ bosons, given by
\begin{equation}\label{fnodeb}
f^{\rm B}_0(\beta,\rho) = \sup_{\mu<0} \left\{ \mu \rho+ \frac n{(2\pi)^3\beta} \int_{\R^3}
dp\, \ln\left( 1 - e^{-\beta (p^2- \mu)}\right)\right\} \,.
\end{equation}
For $\rho<\rho_c(\beta)$, the supremum in (\ref{fnodeb}) is attained at
$\mu=\mu^{\rm B}_0(\beta,\rho)<0$. Denote the fugacity by
$z=e^{\beta\mu}<1$ for this value of $\mu$. Again, $z$ depends only
on the dimensionless quantity $\beta\rho^{2/3}$. Analogously to (\ref{defg0}), let 
\begin{equation}\label{defg0b}
\gamma^{\rm B}_0(p) = \frac 1 {z^{-1} e^{\beta p^2}-1}\,,
\end{equation}
and let $\widetilde\gamma^{\rm B}_0(x)=(2\pi)^{-3} \int dp\, \gamma^{\rm B}_0(p) e^{ipx}$ denote its inverse Fourier transform.

\begin{thm}[High Density Asymptotics for Bosons]\label{T2}
As $\rho\to\infty$ and $\beta\to 0$ (with $\beta\rho^{2/3}<\beta \rho_c(\beta)^{2/3}$),
\begin{equation}\label{thmeqb}
f^{\rm B}(\beta,\rho,\alpha) = f^{\rm B}_0(\beta,\rho) + \frac  {\alpha n}2 \int_{\R^3}dx\,
\frac{|\widetilde\gamma^{\rm B}_0(x)|^2}{|x|} - o(\rho^{4/3}) \,,
\end{equation}
with $0\leq o(\rho^{4/3}) \leq C(\beta\rho^{2/3})\alpha \rho^{4/3}
(\alpha\rho^{-1/3})^{1/48}$. Moreover, the function $C(\beta\rho^{2/3})$ is uniformly bounded on
compact intervals in $(0,\beta\rho_c(\beta)^{2/3})$. 
\end{thm}

As in the fermionic case, the first term on the right side of
(\ref{thmeqb}) is $O(\rho^{5/3})$, whereas the second term is
$O(\rho^{4/3})$. Note that the second term diverges as $\rho\to
\rho_c(\beta)$. This shows that (\ref{thmeqb}) can not hold uniformly
as $\rho$ approaches the critical density, since $f^{\rm
  B}(\beta,\rho,\alpha)\leq f^{\rm B}(\infty,\rho,\alpha)\leq 0$ for
any $\beta$ and $\rho$. At zero temperature, the leading term in the energy
density as $\rho\to\infty$ is actually $O(\rho^{5/4})$
\cite{lcy,ls}. In particular, first order perturbation theory (in the
grand canonical ensemble) is not applicable below the critical
temperature, due to the large fluctuations in particle number. These
large fluctuations cannot be present in the interacting system, for
any non-zero value of the coupling parameter $\alpha$.

The key ingredient in the proof of Theorems~\ref{T1} and~\ref{T2} is a
new correlation estimate, which we present next.

\section{Correlation Estimate}\label{corrsect}

In this section, we will describe our main correlation estimate, which
will then be used in the proof of Theorems~\ref{T1} and~\ref{T2}. For
$\xi \in \R^3$ and $r>0$, let $\chi_{r,\xi}$ denote the
characteristic function of a ball of radius $r$ centered at $\xi$.
The function $\chi_{r,\xi}$ defines a projection operator on
$L^2(\R^3;\C^n)$ and also, in a natural way, on the subspace $L^2(\Lambda;\C^n)$. 
Let $n_{r,\xi}$ denote the operator on Fock space that counts the
number of particles in this ball, i.e., the second quantization of the
projection $\chi_{r,\xi}$ on $\Hh=L^2(\Lambda;\C^n)$.  Our correlation
estimate concerns a lower bound on the expectation value of the number
of pairs of particles inside a ball of radius $r$ or, more precisely,
on
\begin{equation}\label{srie}
\int_{\R^3} d\xi\, \Tr\left[ n_{r,\xi}\left(n_{r,\xi}-1\right) \Gamma\right]\,.
\end{equation}
Here, $\Gamma$ is a density matrix, i.e., a positive operator on Fock
space with trace equal to one, defining the state of the system.

Let $\gamma_0$ denote the one-particle density matrix of a
(grand-canonical) non-interacting (Fermi or Bose) gas at temperature
$T=\beta^{-1}$, with chemical potential $\mu=\mu^{\rm
  F,B}_0(\beta,\rho)$, as defined after Eqs. (\ref{fnode}) and
(\ref{fnodeb}), respectively. We choose {\it periodic} boundary
conditions for $\gamma_0$, which has the advantage of $\gamma_0$
having a constant density. Note that the choice of $\mu$ implies that
$|\Lambda|\bar\rho\equiv \tr\, \gamma_0= |\Lambda|\rho + o(|\Lambda|)$
in the thermodynamic limit. Here and in the following, we denote the
trace over the one-particle space $\Hh=L^2(\Lambda;\C^n)$ by $\tr$,
whereas the trace over Fock space is denoted by $\Tr$. The kernel of
$\gamma_0$ is given by
\begin{equation}\label{gamma0}
\gamma_0(x,\sigma;y,\tau) = \frac 1{|\Lambda|} \sum_{p\in \frac{2\pi}{L}\Z^3}
\gamma_0^{\rm F,B}(p) e^{ip(x-y)} \delta_{\sigma,\tau}\,,
\end{equation}
where $\gamma^{\rm F,B}_0(p)$ is given in
(\ref{defg0}) and (\ref{defg0b}), respectively, and $\sigma$ and $\tau$ label the spin states.

Let $\Gamma_0$ denote the quasi-free state on Fock space with
one-particle density matrix $\gamma_0$. It is the Gibbs state (at
inverse temperature $1$ and chemical potential $0$) for a non-interacting system with
one-particle Hamiltonian $\ln[(1\mp\gamma_0)/\gamma_0]$. Here and in
the following, $\mp$ means $-$ for fermions and $+$ for bosons (and
vice versa for $\pm$). 
We note that for
$\Gamma=\Gamma_0$, the expression in (\ref{srie}) can be
easily calculated. Namely, for any $r$ and $\xi$,

\begin{equation}\label{fixrxi}
\Tr\left[
n_{r,\xi}(n_{r,\xi}-1) \Gamma_0\right] = \left(\tr\left[
\chi_{r,\xi} \gamma_0\right] \right)^2 \mp \tr\left(\chi_{r,\xi}\gamma_0\right)^2\,.
\end{equation}
Hence, after integration over $\xi$, 
\begin{align}\nonumber
  &\int_{\R^3} d\xi \, \Tr\left[ n_{r,\xi}\left(n_{r,\xi}-1\right)
    \Gamma_0\right] \\ &= \int_{\Lambda\times\Lambda} dx\, dy\,
  J_r(x-y) \left[ \bar\rho^2 \mp \mbox{$\sum_\sigma$}
    |\gamma_0(x,\sigma;y,\sigma)|^2\right]\,,\label{intxi}
\end{align}
where we denoted $J_r(x)=\int dy\, \chi_{r,\xi}(y)
\chi_{r,\xi}(x-y)$. (Note that $J_r$ is independent of $\xi$.) Here, we have also
used that $\gamma_0$ has a constant density $\bar\rho=\sum_\sigma
\gamma_0(x,\sigma;x,\sigma)$ for $x\in\Lambda$.

We want to show that for states $\Gamma$ that are in some sense close
to the state $\Gamma_0$, the expectation value (\ref{srie}) is close
to (\ref{intxi}). A convenient way to characterize this ``proximity''
is the {\it relative entropy}:
For two general states $\Gamma$ and $\Upsilon$ on Fock space, the relative entropy is given by
\begin{equation}
S(\Gamma,\Upsilon) =\Tr\, \Gamma ( \ln\Gamma -\ln\Upsilon)\,.
\end{equation}
Note that $0\leq S(\Gamma,\Upsilon)\leq \infty$.  Although $S$ does
not define a metric, it measures the difference between two states in
a certain sense. In particular, $S$ dominates the trace norm. More
precisely, $S(\Gamma,\Upsilon)\geq 2 \|\Gamma-\Upsilon\|_1^2$
\cite[Thm.~1.15]{ohya}.

Note that the relative entropy can also be interpreted as a difference
in free energies. More precisely, if $\Upsilon=\exp(-\beta(H-F))$ for some $\beta>0$, with
$F=-\beta^{-1} \ln \Tr \exp(-\beta H)$ the corresponding \lq\lq free energy\rq\rq, then
\begin{equation}\label{fre}
\beta^{-1} S(\Gamma,\Upsilon)= \Tr [H\Gamma] + \beta^{-1} \Tr\, \Gamma\ln\Gamma - F\,.
\end{equation}
Note that $-\Tr\, \Gamma\ln\Gamma$ is just the von-Neumann
entropy of $\Gamma$. Hence the first two terms on the right side of (\ref{fre})
correspond to the free energy of $\Gamma$ (with Hamiltonian and
temperature determined by $\Upsilon$), whereas $F$ is the free energy of $\Upsilon$.

Our main result estimates the difference of the expectation value
(\ref{srie}) for $\Gamma_0$ and a general state $\Gamma$ in terms of
the relative entropy $S(\Gamma,\Gamma_0)$.  More precisely, the
following Theorem, which is the main new result of this work, holds.

\begin{thm}[Main Correlation Estimate]\label{P1}
Let $\Gamma_0$ be given as above, with one-particle density matrix
$\gamma_0$ and density $\bar\rho$, and with $\mu\in \R$ for fermions and
$\mu<0$ for bosons. Let $\Gamma$ be any other state
on (fermionic or bosonic) Fock space. For any $2r \leq d \leq L/2$, we have that
\begin{align}\nonumber 
&\int_{\R^3} d\xi \, \Tr\left[ n_{r,\xi}\left(n_{r,\xi}-1\right)
\Gamma\right] \\ &\geq  \int_{\Lambda\times\Lambda} dx\, dy\, J_r(x-y)
\left[ \bar\rho^2 \mp \mbox{$\sum_\sigma$} \label{t3e}
  |\gamma_0(x,\sigma;y,\sigma)|^2\right]  \\ & \quad -C^{\rm F,B}_z r^3\bar\rho
\left(1+ r^3\bar\rho\right)|\Lambda|^{3/4}\left[ d^{3}\left(1+\beta d^{-2}\right)
  S(\Gamma,\Gamma_0)+  \beta^{1/2} d^{-1}|\Lambda|\right]^{1/4}\,. \nonumber
\end{align}
Here, $C^{\rm F,B}_z$ are constants depending only on
$z=e^{\beta\mu}$, which are uniformly bounded on compact
intervals in $(0,\infty)$ and $(0,1)$, respectively. 
\end{thm}

We emphasize again that, according to (\ref{intxi}), the second line
in (\ref{t3e}) equals the first in the case
$\Gamma=\Gamma_0$. Although the inequality (\ref{t3e}) is not sharp in
this case, the parameter $d$ can be made very large to obtain an error
with is, in the thermodynamic limit, of lower order than the volume. (The restriction $d\leq L/2$ in
Theorem~\ref{P1} is purely technical and could in principle be avoided
by a slight modification of the proof. Since we are mainly concerned
here with the application of (\ref{t3e})
in the thermodynamic limit $L\to \infty$, we have refrained
from doing so.) 

Note that Theorem~\ref{P1} gives an estimate on a ``local'' quantity,
like the expectation value of the number of pairs of particles inside
a small ball, in terms of a ``global'' quantity as the relative
entropy. The {\it strong subadditivity} of entropy plays a crucial role in
this estimate. Before we give the proof of Theorem~\ref{P1}, we show how
it can be used to prove the applications to Coulomb systems stated in
Theorems~\ref{T1} and~\ref{T2}.

\section{Proof of Theorems~\ref{T1} and~\ref{T2}}\label{lowsect}

We are going to treat the fermionic and bosonic case simultaneously,
merely pointing out the differences if necessary.  We start by
deriving a lower bound on the free energy. Note that if $\Gamma$
denotes the Gibbs state of $H$ at temperature $\beta^{-1}$ (and zero
chemical potential), then charge neutrality (as proved in
\cite{liebnarn}) implies that
\begin{equation}
\lim_{L\to\infty} \frac 1{|\Lambda|} \Tr\, N \Gamma = \rho
\end{equation}
for any fixed $\beta$ and $\alpha>0$. Here, $N$
denotes the number operator on Fock space. 
Application of the Peierls-Bogoliubov inequality then leads to the lower bound 
\begin{equation}\label{42i}
f^{\rm F,B}(\beta,\rho,\alpha)\geq f^{\rm F,B}_0(\beta,\rho) + \alpha \limsup_{L\to
  \infty} \frac 1{|\Lambda|} \Tr\, W \Gamma\,.
\end{equation}
To estimate the expectation value of $W$ in the Gibbs state $\Gamma$,
we will split the Coulomb potential into a long and short-range part.

\subsection{Long-Range Part}
We write the Coulomb potential as \cite{feff}
\begin{equation}\label{feffl}
\frac{1}{|x-y|} =\frac 1\pi   \int_0^\infty dr\, \frac 1{r^5}\,  \int_{\R^3} d\xi\, 
\chi_{r,\xi}(x)\chi_{r,\xi}(y)\,.
\end{equation}
As in Section~\ref{corrsect}, $\chi_{r,\xi}$ denotes the characteristic function of a ball of
radius $r$ centered at $\xi\in\R^3$. We split the $r$-integration into a part
$r\leq R$ and a part $r\geq R$ and, correspondingly, write
\begin{equation}
\frac{1}{|x-y|} = V_{<R}(x-y) + V_{>R}(x-y)\,.
\end{equation}
Note that $V_{<R}(x)=0$ for $|x|\geq 2R$. For the long-range part
$V_{>R}$, we note that it has a positive Fourier transform, as follows
immediately from the decomposition (\ref{feffl}). Hence we obtain the lower bound \cite[4.5.20]{thirring}
\begin{align}\label{comb1}
\sum_{1\leq i<j\leq N} V_{>R}(x_i-x_j) \geq & \sum_{i=1}^N \rho\int_\Lambda
dy\, V_{>R}(x_i-y) \\ & - \half \rho^2 \int_{\Lambda\times\Lambda} dy_1\, dy_2\,
V_{>R}(y_1-y_2) - \frac N2 V_{>R}(0)\,. \nonumber
\end{align}
This estimate actually holds for
any $\rho>0$. 
The last term equals $V_{>R}(0)=4/(3R)$, and hence will be negligible if we
choose $R\gg \rho^{-1/3}$.

\subsection{Short-Range Part}
As in Section~\ref{corrsect}, let $n_{r,\xi}$ denote the operator that counts the
number of particles in a ball of radius $r$ centered at $\xi$, i.e., the
second quantization of the projection $\chi_{r,\xi}$ on $\Hh=L^2(\R^3;\C^n)$. The
expectation value of the short-range part $V_{<R}$ of the
interparticle interaction in a state $\Gamma$ on Fock space
can be written as 
\begin{equation}\label{srie1}
\frac 1{2\pi}\int_0^R dr\, \frac 1{r^5}\, \int_{\R^3} d\xi \, \Tr\left[ n_{r,\xi}\left(n_{r,\xi}-1\right)
\Gamma\right]\,.
\end{equation}
For a lower bound, we can now apply our main correlation estimate,
Theorem~\ref{P1}, to the expression (\ref{srie1}) for any fixed $r$.

Recall that $\gamma_0$ denotes the one-particle density matrix of a
non-inter\-acting (Fermi or Bose) gas at inverse temperature
$\beta$, with chemical potential $\mu=\mu^{\rm F,B}_0(\beta,\rho)$, and
with periodic boundary conditions; $\Gamma_0$ denotes the
corresponding quasi-free state on Fock space. Theorem~\ref{P1} states that 
for any $2r \leq d \leq L/2$,
\begin{align}\nonumber 
&\int_{\R^3} d\xi \, \Tr\left[ n_{r,\xi}\left(n_{r,\xi}-1\right)
\Gamma\right] \\ & \label{propim} \geq  \int_{\Lambda\times\Lambda} dx\, dy\, J_r(x-y)
\left[ \bar\rho^2 \mp \mbox{$\sum_\sigma$}
  |\gamma_0(x,\sigma;y,\sigma)|^2\right]  \\ & \quad -C^{\rm F,B}_z r^3\bar\rho
\left(1+ r^3\bar\rho\right)|\Lambda|^{3/4}\left[ d^{3}\left(1+\beta d^{-2}\right)
  S(\Gamma,\Gamma_0)+  \beta^{1/2} d^{-1}|\Lambda|\right]^{1/4}\,. \nonumber
\end{align}
For $\Gamma$ the Gibbs state of $H$, an upper bound on
$S(\Gamma,\Gamma_0)$ is, in fact, easy to obtain. Using the fact that
the quadratic form domain of the Dirichlet Laplacian is contained in
the quadratic form domain of the Laplacian on $\Lambda$ with periodic
boundary conditions, we can write
\begin{align}\nonumber
S(\Gamma,\Gamma_0) = & \beta\, \Tr\, (H_0 - \mu N) \Gamma + \Tr\,
\Gamma\ln\Gamma \mp \tr\, \ln(1\mp \gamma_0) \\  = & - \ln \Tr\, \exp[-\beta
H] - \beta\mu\, \Tr\, N\Gamma - \beta \alpha\, \Tr\, W\Gamma \mp \tr\,
\ln(1\mp\gamma_0)\,.
\end{align}
We now use the lower bound $W\geq
- \const N |\Lambda| \rho ^{1/3}$ \cite{liebnarn}, as well as the fact
that $|\Lambda|^{-1}\Tr\, N \Gamma \to \rho$ in the thermodynamic
limit, as explained in the beginning of this section. This leads to the estimate
\begin{equation}\label{gg0a}
S(\Gamma,\Gamma_0) \leq |\Lambda| \beta \left( f^{\rm
    F,B}(\beta,\rho,\alpha)-f^{\rm F,B}_0(\beta,\rho) \right) + \const
\beta |\Lambda| \alpha \rho^{4/3} + o(|\Lambda|)\,.
\end{equation}
As the upper bound to the free energy in Section~\ref{upsect} shows,
the first term is negative in the fermionic case and can thus be
neglected for an upper bound. In the bosonic case, it is bounded above
by $C_z \beta |\Lambda| \alpha \rho^{4/3}$ for some constant depending
only on $z$. This follows immediately from the upper bound leading to
(\ref{thmeqb}), together with simple
scaling. (Note that $C_z$ diverges as $z\to 1$). Hence, in general,
\begin{equation}\label{gg0}
S(\Gamma,\Gamma_0) \leq C_z \beta |\Lambda| \alpha \rho^{4/3} +
o(|\Lambda|)\,.
\end{equation}
Here and in the following, we abuse the notation slightly and denote
by $C_z$ any expression that depends only on $z$ (and is uniformly
bounded on compact intervals in $(0,\infty)$ in the fermionic case and
$(0,1)$ in the bosonic case).

We insert the bound (\ref{gg0}) into (\ref{propim}).
Choosing $d=\beta^{-1/8} \alpha^{-1/4}\rho^{-1/3}$ we thus obtain that, as long as $d\geq 2r$ (and
$\alpha\rho^{-1/3}\leq \const$),
\begin{align}\nonumber 
& \int_{\R^3} d\xi \, \Tr\left[ n_{r,\xi}\left(n_{r,\xi}-1\right)
\Gamma\right] \\ \nonumber & \geq  \int_{\Lambda\times\Lambda} dx\, dy\, J_r(x-y)
\left[ \rho^2 \mp \mbox{$\sum_\sigma$}
  |\gamma_0(x,\sigma;y,\sigma)|^2\right]  \\& \quad -C_z r^3\rho \left(1+
  r^3\rho\right)|\Lambda| \left(\beta^{5/2} \alpha
  \rho^{4/3}\right)^{1/16} +o(|\Lambda|)\,. \label{rdl}
\end{align}
Here, we have also used that $\bar\rho=\rho+o(1)$ as $L\to\infty$. 
Note that we are going to use this estimate in (\ref{srie1}) only for $r\leq R$. 
Below we will choose $R\ll \beta^{-1/8}
\alpha^{-1/4}\rho^{-1/3}$, hence (\ref{rdl}) will be applicable.

For a lower bound, we can restrict the $r$-integration in
(\ref{srie1}) to $R_0\leq r\leq R$ for some $0<R_0<R$, and simply
neglect the contribution from the $r\leq R_0$ part. A simple estimate,
using that $ \mbox{$\sum_\sigma$} |\gamma_0(x,\sigma;y,\sigma)|^2 \leq
\bar\rho^2$, shows that in the state $\Gamma_0$ this $r\leq R_0$
contribution is bounded above by $(2\pi)^{-1} |\Lambda|
(4\pi/3)^2 \bar\rho^2 R_0^2$.  We then have
\begin{align}\nonumber
&\frac 1{2\pi} \int_0^R dr\, \frac 1{r^5} \int_{\R^3} d\xi\, \Tr\left[ n_{r,\xi}\left(n_{r,\xi}-1\right)
\Gamma\right] \\ \nonumber &\geq \frac 12 \int_{\Lambda\times\Lambda} dx\, dy\,
V_{<R}(x-y) \left[ \rho^2 \mp \mbox{$\sum_\sigma$}
  |\gamma_0(x,\sigma;y,\sigma)|^2\right] - |\Lambda|
\frac{8\pi}9 \rho^2 R_0^2 \\ \ & \quad
- C_z \rho |\Lambda| \left(\frac 1{R_0}+ R^2\rho\right)  \big(\alpha
  \rho^{-1/3}\big)^{1/16}+o(|\Lambda|)\,. \label{comb3}
\end{align}
Here, we have used again that $z$ is a function of $\beta\rho^{2/3}$. 

\subsection{Final Lower Bound}
For the one-body part containing $V_{<R}$ (i.e., the interaction with
the background), we can use the simple lower
bound 
\begin{equation}\label{comb4}
-\sum_{i=1}^N \rho \int_{\Lambda}dx\, V_{<R}(x-x_i) \geq - N \rho \int_{\R^3} dx\, V_{<R}(x)\,.
\end{equation}
In combination, (\ref{comb1}), (\ref{comb3}) and (\ref{comb4}) yield,
in the thermodynamic limit,
\begin{align}\nonumber
\liminf_{L\to\infty} \frac 1{|\Lambda|} \Tr\, W \Gamma \geq
&\mp \frac n2 \int_{\R^3} dx\, V_{<R}(x)|\widetilde\gamma^{\rm F,B}_0(x)|^2 - \frac {2\rho}{3R}-  
\frac{8\pi}9 \rho^2 R_0^2 \\ &- C_z \rho \left(\frac 1{R_0}+  R^2\rho\right) \big(\alpha\rho^{-1/3}\big)^{1/16}\,.
\end{align}
In the fermionic case, we can simply use $V_{<R}(x)\leq |x|^{-1}$ for a
lower bound. In the bosonic case, we write $V_{<R}(x)= |x|^{-1} -
V_{>R}(x)$ and estimate (using $V_{>R}(x)\leq V_{>R}(0)=4/(3R)$)
\begin{equation}
\frac n2 \int_{\R^3} dx\, V_{>R}(x)|\widetilde\gamma^{\rm B}_0(x)|^2 \leq \frac
{2n}{3R}(2\pi)^{-3} \int_{\R^3} dp\, |\gamma^{\rm B}_0(p)|^2 \leq \frac
{2\rho}{3R} \frac {z}{1-z}\,. 
\end{equation}
With the choice
$R=\rho^{-1/3}(\alpha\rho^{-1/3})^{-1/48}$ and $R_0=1/(R^2\rho)$ this yields
\begin{equation}
\liminf_{L\to\infty} \frac 1{|\Lambda|} \Tr\, W \Gamma \geq
\mp \frac n2 \int_{\R^3}\frac{|\widetilde\gamma^{\rm
    F,B}_0(x)|^2}{|x|} 
- C_z  \rho^{4/3} \, 
\big(\alpha\rho^{-1/3}\big)^{1/48}
\end{equation}
for some constant $C_z$ depending only on $z$. Inserting this bound into (\ref{42i}) finishes the proof
of the lower bound.

\subsection{Upper Bound}\label{upsect}
For the upper bound to the free energy, we use the variational principle, which states that  
\begin{equation}
-\frac 1\beta \ln \Tr\,\exp[-\beta H] \leq  \Tr\, H \Gamma - \frac
1\beta S(\Gamma)
\end{equation}
for any state $\Gamma$ on Fock space. Here, $S(\Gamma)=-\Tr\,
\Gamma\ln\Gamma$ denotes the von-Neumann entropy. We choose as a
trial state $\Gamma$ a quasi-free state with one-particle density matrix $\gamma$ given
by the kernel
\begin{equation}\label{opdm}
\gamma(x,\sigma;y,\tau) = g(x) g(y) \widetilde\gamma^{\rm F,B}_0(x-y)
\delta_{\sigma,\tau}\,.
\end{equation}
Here, $0\leq g(x)\leq 1$ is a continuously differentiable function with the property that $g(x)=0$ for
$x\not\in \Lambda$, $g(x)=1$ if $x\in\Lambda$ and ${\rm
  dist}(x,\partial\Lambda)\geq R$, and $|\nabla g|\leq \const
R^{-1}$. We shall choose the variational parameter  $R$ to satisfy
$1/(L\rho^{2/3})\ll R \ll (L\rho^2)^{-1/5}$ for large $L$.

The calculation of the energy of the state $\Gamma$ is similar to the
corresponding calculation in \cite{grafsol}. It is in fact simpler
since the particle number does not have to be fixed. 

Let $\varphi(x)= \sum_{\sigma}\gamma(x,\sigma;x,\sigma)$ denote the
density of $\gamma$. A simple computation (compare with (\ref{fixrxi})--(\ref{intxi})), using the fact that $\Gamma$
is a quasi-free state, yields
\begin{equation}\label{419}
\Tr\, W\Gamma = \half \int_{\Lambda\times\Lambda} dx\, dy\,
\frac{1}{|x-y|}\left[ (\varphi(x)-\rho)(\varphi(y)-\rho)  \mp \mbox{$\sum_{\sigma}$}|\gamma(x,\sigma;y,\sigma)|^2\right]\,.  
\end{equation}
Note that $\varphi(x)=\rho(1-g(x)^2)$ and hence, by definition, $\varphi(x)=\rho$ if $x\in \Lambda$ and
$x$ is at least a distance $R$ away from the boundary of $\Lambda$.
Using the Hardy-Littlewood-Sobolev inequality
\cite[Thm.~4.3]{liebloss}, it is easy to see that the first term on the right side of (\ref{419}) is
bounded from above by $\const \rho^2 (L^2 R)^{5/3}$ and is thus
negligible in the thermodynamic limit, if $R\ll (L\rho^2)^{-1/5}$.
In the fermionic case, the second term is bounded from above by
\begin{equation}
- \frac n2 \int_{\Lambda\times\Lambda} dx\, dy\,
\frac{1}{|x-y|} |\widetilde \gamma^{\rm F}_0(x-y)|^2 \left(1 - 2
  (1-g(x)^2)\right)\,,
\end{equation}
which yields the desired expression in the thermodynamic
limit, provided $R\ll L$, which is amply satisfied for our choice of
$R$. In the bosonic case, we can simply use $g\leq 1$ to obtain the
desired bound.

The kinetic energy of $\Gamma$ is given by
\begin{align}\nonumber
\Tr\, H_0 \Gamma& = - n \Delta \widetilde\gamma^{\rm F,B}_0(0) \int_{\R^3} dx\,
g(x)^2  + \rho \int_{\R^3} dx\, |\nabla g(x)|^2 \\ &\leq -n |\Lambda|
\Delta\widetilde\gamma^{\rm F,B}_0(0) + \const \frac {\rho L^2}{R}\,. \label{enup}
\end{align}
Again, the first term is the desired expression, and the last term is
negligible if $R\gg 1/(L\rho^{2/3})$.

It remains to derive a lower bound on the entropy $S(\Gamma)$.  
We claim that 
\begin{align}\nonumber
S(\Gamma) \geq & 
-\frac n{(2\pi)^{3}} \int_{\Lambda} dx\, g(x)^2 \\ & \times  \int_{\R^3}dp\,
\left[ \gamma^{\rm F,B}_0(p) \ln \gamma^{\rm F,B}_0(p)
\pm \left(1\mp \gamma^{\rm F,B}_0(p)\right) 
\ln\left(1\mp\gamma^{\rm F,B}_0(p)\right)\right]\,, \label{bl}
\end{align}
which gives the desired quantity as long as $R\ll L$. Inequality
(\ref{bl}) follows from a variant of the Berezin-Lieb inequality
\cite{berezin,liebb}. The one-particle density matrix (\ref{opdm}) can
be written as
\begin{equation}
\gamma = \sum_{\sigma} \int_{\R^3} dp\, \gamma_0^{\rm F,B}(p) g
|p,\sigma\rangle\langle p,\sigma| g\,,
\end{equation}
where $g$ denotes multiplication by $g(x)$, and $|p,\sigma\rangle$
denotes a plane wave with wave function $(2\pi)^{-3/2} \exp(ipx)$ and
spin $\sigma$. Moreover, since $\Gamma$ is a quasi-free state,
\begin{equation}
S(\Gamma) = \tr\, s(\gamma)\,,
\end{equation}
where we denoted $s(t) = -t\ln t \mp (1\mp t)\ln(1\mp t)$ for $t\geq
0$. Note that $s$ is a concave function, with $s(0)=0$. Hence we can 
apply the Berezin-Lieb inequality, in the form proved in Thm.~A1 in
\cite{solo}. Noting that 
\begin{equation}
 \sum_{\sigma} \int_{\R^3} dp\,  g
|p,\sigma\rangle\langle p,\sigma| g = g^2 \leq 1\,,
\end{equation}
as well as $\langle p,\sigma|g^2|p,\sigma\rangle = (2\pi)^{-3} \int
dx\, g(x)^2$, this yields (\ref{bl}).

We conclude that
\begin{align}\nonumber
- &\lim_{L\to\infty}  \frac{1}{\beta|\Lambda|} \ln\Tr\, \exp[-\beta
H] \leq  \mp \frac {n\alpha}2 \int_{\R^3} dx\,
\frac {|\widetilde \gamma^{\rm F,B}_0(x)|^2}{|x|} -n
\Delta\widetilde \gamma^{\rm F,B}_0(0) \\ &+\frac n{(2\pi)^{3}\beta}
\int_{\R^3}dp\,\left[ \gamma^{\rm F,B}_0(p) 
\ln \gamma^{\rm F,B}_0(p)
\pm \left(1\mp\gamma^{\rm F,B}_0(p)\right) 
\ln\left(1\mp\gamma^{\rm F,B}_0(p)\right)\right]\,.
\end{align}
The last two terms together are just $f^{\rm F,B}_0(\beta,\rho)$. 
We have thus established the desired upper bounds. This concludes the proof of Theorems~\ref{T1} and~\ref{T2}.

\section{Proof of Theorem~\ref{P1}}

\subsection{Localization of Relative Entropy}

If $X$ denotes a projection on the one-particle space
$\Hh$, then states on the Fock space can be restricted
to the Fock space over the subspace $X\Hh$ of $\Hh$. We denote
such a restriction of a state $\Gamma$ by $\Gamma_X$. Since $\chi_{r,\xi}$
defines a projection on $\Hh=L^2(\Lambda;\C^n)$, we can write
\begin{equation}
 \Tr\left[ n_{r,\xi}\left(n_{r,\xi}-1\right)
\Gamma\right]= \Tr\left[ n_{r,\xi}\left(n_{r,\xi}-1\right)
\Gamma_{\chi_{r,\xi}}\right]\,,
\end{equation}
the latter trace being over the Fock space over $\chi_{r,\xi}\Hh$. 

It is well known \cite{ohya} that
the relative entropy decreases under restriction. More precisely,
for any two states $\Gamma$ and $\Upsilon$ on Fock space, 
\begin{equation}
S(\Gamma,\Upsilon) \geq S(\Gamma_{X},\Upsilon_{X})\,.
\end{equation}
This property is closely related to the strong subadditivity of the von-Neumann entropy
\cite{liebrus,lind}.

Let $\eta:\R^3\mapsto\R$ be a function with the following properties:
\begin{itemize}
\item $\eta\in C^4(\R^3)$
\item $\eta(0)=1$, and $\eta(x)=0$ for $|x|\geq 1$
\item $\widehat \eta (p)=\int dx\, \eta(x) e^{-ipx} \geq 0$ for all $p\in\R^3$.
\end{itemize}
We note that such a function (with any degree of regularity) can, for
instance, be obtained
by taking a smooth function of compact support, and convolving it with
itself. The resulting function  is then smooth, has compact support and
positive Fourier transform. In our application, we need the existence
of the fourth derivatives at the origin (see Eq. (\ref{4der}) below). 

Given such a function $\eta$, we define $\eta_d(x)=\eta(x/d)$ and
\begin{equation}
\eta_d^{\per}(x)=\sum_{j\in \Z^3} \eta_d(x+ j L)\,.
\end{equation}
Note that $\eta_d^\per$ is a periodic function with period $L$ and, since $L\geq 2d$ by assumption, we have that $\eta_d^\per\leq
1$. 
 Moreover, we define a one-particle density
matrix $\gamma_d$ on $\Hh$ by the kernel
\begin{equation}\label{defgd}
\gamma_d(x,\sigma;y,\tau)=\gamma_0(x,\sigma;y,\tau) \eta_d^\per(x-y)\,,
\end{equation}
with $\gamma_0$ defined in (\ref{gamma0}). 
This defines a positive
operator, with plane waves as eigenfunctions, and eigenvalues
determined by the convolution of $\widehat \eta_d$ and $\gamma^{\rm F,B}_0(p)$.

If $[L/2d]$ denotes the largest integer $\leq L/2d$, define $\bar d$
by $L/2\bar d = [L/2d]$. Then $d\leq \bar d \leq 2d$.
For $0\leq r \leq d/2$, let $X_r$ denote the characteristic function
of a collection of balls of radius $r$, separated by $2\bar d$:
\begin{equation}
X_r(x)=\sum_{\xi\in 2\bar d\,\Z^3} \chi_{r,\xi}(x) = \sum_{\xi\in
  2\bar d\,\Z^3 \cap [0,L)^3 } \chi^\per_{r,\xi}(x)\,,
\end{equation}
where we denoted 
\begin{equation}
 \chi^\per_{r,\xi}(x) = \sum_{j\in \Z^3} \chi_{r,\xi}(x+jL)\,.
\end{equation}
Note that the minimal distance between the balls is $2\bar d-2r\geq
d$. Hence
\begin{equation}\label{xgdx}
X_r\, \gamma_d\, X_r = \sum_{\xi\in 2\bar d\,\Z^3\cap [0,L)^3} \chi^\per_{r,\xi} \, \gamma_d \,
\chi^\per_{r,\xi}\,,
\end{equation}
the off-diagonal terms vanish since $\eta_d(x)=0$ for $|x|\geq
d$. I.e., $X_r\gamma_dX_r$ is a direct sum of one-particle density
matrices on $\chi^\per_{r,\xi}\Hh$ for $\xi\in 2\bar d\Z^3\cap [0,L)^3$.

Let $\Gamma_d$ denote the quasi-free state on Fock space with
one-particle density matrix $\gamma_d$, and let $\Gamma$ denote any
other state on Fock space.  The characteristic function $X_r$ defines
a projection operator on the one-particle space $\Hh=L^2(\Lambda;\C^n)$.
Hence the monotonicity of the relative entropy implies 
\begin{equation}
S(\Gamma,\Gamma_d) \geq S(\Gamma_{X_r},\Gamma_{d,X_r})\,,
\end{equation}
where $\Gamma_{X_r}$ and $\Gamma_{d,X_r}$ denote the states restricted
to the Fock space over $X_r\Hh$, respectively. Note that the
one-particle density matrix of the quasi-free state $\Gamma_{d,X_r}$
is given by $X_r \gamma_d X_r$. Hence (\ref{xgdx}) shows that
$\Gamma_{d,X_r}$ can be written as a product of states on the Fock
spaces over the one-particle spaces $\chi^\per_{r,\xi}\Hh$ for $\xi\in
2\bar d\,
\Z^3\cap [0,L)^3$. Under this condition $S$ is superadditive, as follows easily from subadditivity of the von-Neumann entropy  \cite{ohya}. More precisely, 
\begin{equation}\label{supera}
S(\Gamma,\Gamma_d) \geq S(\Gamma_{X_r},\Gamma_{d,X_r}) \geq \sum_{\xi\in
  2\bar d\, \Z^3\cap [0,L)^3} S(\Gamma_{\chi^\per_{r,\xi}},\Gamma_{d,\chi^\per_{r,\xi}})\,.
\end{equation}

We can repeat the argument above with a projector defined by the
multiplication operator $X_r(x+a)$ for some vector $a\in [0,2\bar d]^3$.
Averaging over $a$ then yields
\begin{align}\nonumber
S(\Gamma,\Gamma_d) &\geq \frac 1{(2\bar d)^3} \int_{[0,2\bar d]^3} da\,
\sum_{\xi\in 2\bar d\, \Z^3\cap [0,L)^3} S(\Gamma_{\chi^\per_{r,\xi+a}},\Gamma_{d,\chi^\per_{r,\xi+a}})
\\ &= \frac 1{(2\bar d)^3}
\int_{\Lambda} d\xi\, S(\Gamma_{\chi^\per_{r,\xi}},\Gamma_{d,\chi^\per_{r,\xi}})\,. \label{hest}
\end{align}

\bigskip
{\it Remark.} We emphasize that in order to obtain the
superadditivity of the relative entropy leading to (\ref{supera}), we
have used the fact that $\Gamma_{d,X_r}= \bigotimes_\xi
\Gamma_{d,\chi^\per_{r,\xi}}$. Our estimate applies to any density
matrix having this property. For a general state, however, it will be
difficult to check this property; in the case of a quasi-free state
considered here, it simply translates to the vanishing of off-diagonal
terms in the one-particle density matrix (more precisely, the validity
of (\ref{xgdx})).

\subsection{Upper Bound on Relative Entropy with Cutoff}

In the previous subsection, we have shown how to localize relative
entropy in the case when the second argument is a state that has been
cut off in such away as to avoid correlations between balls of a
certain distance. In the following, we will quantify the effect of this
cut-off on the relative entropy. 

If $\Gamma_\gamma$ denotes the quasi-free state with one-particle density matrix
$\gamma$, and $\Upsilon$ is any other state on Fock space, then
$S(\Upsilon,\Gamma_\gamma)$ is convex in $\gamma$. This follows from
operator-concavity of the logarithm and 
$S(\Upsilon,\Gamma_\gamma)=\Tr\, \Upsilon\ln\Upsilon - \tr\, \omega \ln
\gamma \mp \tr\, (1\mp\omega)\ln(1\mp\gamma)$, where $\omega$ denotes the
one-particle density matrix of $\Upsilon$. Note that
$\gamma_d$ can be written as a convex combination of the form
\begin{equation}
\gamma_d = \frac{1}{|\Lambda|} \sum_{q\in \frac{2\pi}L \Z^3}
\widehat \eta_d(q) \, \gamma_{0,q}
\end{equation}
where $\gamma_{0,q}$ is defined by the kernel 
\begin{equation}
\gamma_{0,q}(x,\sigma;y,\tau) = \frac 1{|\Lambda|} \sum_{p\in \frac{2\pi}L \Z^3}
\frac12\left[\gamma^{\rm F,B}_0(p+q)+\gamma^{\rm F,B}_0(p-q)\right] e^{i p (x-y)} \delta_{\sigma,\tau}\,.
\end{equation}
Hence convexity implies that, for any state $\Gamma$, 
\begin{equation}\label{qsum}
S(\Gamma,\Gamma_d)\leq \frac{1}{|\Lambda|} \sum_{q\in \frac{2\pi}L
  \Z^3}\widehat \eta_d(q) \, S(\Gamma,\Gamma_{0,q})\,,
\end{equation}
where $\Gamma_{0,q}$ denotes the quasi-free state corresponding to the one-particle density matrix 
$\gamma_{0,q}$. 

Recall that $\Gamma_0$ denotes the quasi-free state on Fock space with
one-particle density matrix $\gamma_0$ given in (\ref{gamma0}), i.e., 
$\Gamma_0\equiv \Gamma_{0,0}$. We
claim that, for any $t>0$,
\begin{equation}\label{3terms}
S(\Gamma,\Gamma_{0,q})\leq \left(1+t^{-1}\right) S(\Gamma,\Gamma_0) +\tr\, (h_q -
h_0)\left(\frac 1{e^{(1+t)h_0 - t h_q}\pm 1} - \frac 1{e^{h_q}\pm 1}\right),
\end{equation}
where $h_q = \ln[(1\mp \gamma_{0,q})/\gamma_{0,q}]$. In the Bose case,
we have to assume that $(1+t)h_0-t h_q> 0$, which is satisfied
for $t$ small enough, as our estimates in Lemma~\ref{L2} below will show.
Inequality (\ref{3terms}) follows from
the two inequalities (where $\gamma$ denotes the one-particle density
matrix of $\Gamma$)
\begin{align}\label{1ine}
& \tr\, \gamma((1+t)h_0-th_q) +\Tr\, \Gamma\ln\Gamma \\ \nonumber &\geq \mp \tr\,
\ln \left(1\pm e^{-(1+t) h_0+th_q}\right)
\\ &\geq \mp \tr\, \ln \left(1\pm e^{-h_0}\right) + t\,\tr\,
(h_0-h_q)[e^{(1+t)h_0-th_q}\pm 1]^{-1} \nonumber
\end{align}
and
\begin{equation}\label{2ine}
\mp \tr\, \ln \left(1\pm e^{-h_q}\right)
\geq \mp \tr\, \ln\left(1\pm e^{-h_0}\right) +  \tr\,
(h_q-h_0)[e^{h_q}\pm 1]^{-1}\,. 
\end{equation}
Dividing  (\ref{1ine}) by $t$  and adding (\ref{2ine})  yields (\ref{3terms}).

To estimate the last term in (\ref{3terms}), we need the
following simple lemmas, estimating the expression 
\begin{equation}
h^{\rm F,B}_q(p) =
\ln\frac{2\mp\gamma^{\rm F,B}_{0}(p+q)\mp\gamma^{\rm F,B}_{0}(p-q)}{\gamma^{\rm F,B}_{0}(p+q)+\gamma^{\rm F,B}_{0}(p-q)}\,.
\end{equation}
Note that $h^{\rm F,B}_0(p) = \beta(p^2-\mu)$. 

\begin{lem}[Fermions]\label{L1}
 Let $D_z=\sup_{u>0}
[zu/(e^u+z)]$.  Then
\begin{equation}\label{q2cz}
-2 \beta q^2(3 D_z + 2\beta p^2) \leq h^{\rm F}_q(p)-h^{\rm F}_0(p) \leq 2 \beta q^2 (1+2D_z) \,.
\end{equation}
Moreover,
\begin{equation}\label{q3cz}
\beta( q^2 - 2 |pq|) \leq h^{\rm F}_q(p)-h^{\rm F}_0(p)\leq \beta( q^2 + 2 |pq|)
\end{equation}
independently of $z$.
\end{lem}

\begin{lem}[Bosons]\label{L2}
 Let $D_z=\sup_{u>0}
[z^2 u e^u/(e^u-z)^2]$.  Then
\begin{equation}\label{q2czb}
-2 \beta q^2\left(3 D_z + 2\beta p^2\right) \leq h^{\rm
  B}_q(p)-h^{\rm B}_0(p) \leq \beta q^2  \,.
\end{equation}
Moreover,
\begin{equation}\label{q3czb}
h^{\rm B}_q(p)-h^{\rm B}_0(p)\geq \beta( q^2 - 2 |pq|) 
\end{equation}
independently of $z$.
\end{lem}

We defer the proof of Lemmas~\ref{L1} and~\ref{L2} to the appendix.

The last term in (\ref{3terms}) is given by
\begin{equation}\label{sumpq4}
n \sum_{p\in \frac{2\pi}L \Z^3}  \left(h^{\rm F,B}_q(p)-h^{\rm F,B}_0(p) \right)
\left( \frac 1{e^{(1+t) h^{\rm F,B}_0(p)-th^{\rm F,B}_q(p)}\pm 1} - 
\frac 1{e^{h^{\rm F,B}_q(p)}\pm 1}
 \right)\,.
\end{equation}
A simple estimate on the derivative of the last term in brackets with
respect to $h^{\rm F,B}_0(p)-h^{\rm F,B}_q(p)$ shows that
(\ref{sumpq4}) is bounded above by 
\begin{equation}\label{sum2f}
%(\ref{sumpq4})\leq 
n (1+t) C_z \sum_{p\in \frac{2\pi}L \Z^3} 
\left(h^{\rm F,B}_q(p)-h^{\rm F,B}_0(p) \right)^2 \sup_{-1\leq s\leq
  t} \frac 1{e^{(1+s)h^{\rm
      F,B}_0(p)-s h^{\rm F,B}_q(p)}\pm 1}\,,
\end{equation}
where $C_z=1$ for fermions and $C_z=(1-z)^{-1}$ for bosons. (Here we
have used that $1+\gamma_0^{\rm B}(p)\leq (1-z)^{-1}$.)  The upper
bounds in (\ref{q2cz}) and (\ref{q2czb}) show that, for $0\leq s\leq
t$,
\begin{equation}\label{lb1}
(1+s)h^{\rm
      F,B}_0(p)-s h^{\rm F,B}_q(p) \geq \left\{ \begin{array}{ll}
        h_0^{\rm F}(p) - 2 t\beta q^2 (1+2D_z) & {\rm for\
          fermions\,,} \\ 
        h_0^{\rm B}(p) - t \beta q^2 & {\rm for\ bosons\,.}
      \end{array}\right. 
\end{equation}
We choose $t=\min\{1,(2\beta q^2(1+2D_z))^{-1}\}$ in the fermionic case,
and $t=\min\{1,-\mu/(2q^2)\}$ in the bosonic case. With this choice,
(\ref{lb1}) becomes
\begin{equation}\label{lb2}
(1+s)h^{\rm
      F,B}_0(p)-s h^{\rm F,B}_q(p) \geq \left\{ \begin{array}{ll}
        \beta(p^2-\mu) -1  & {\rm for\
          fermions\,,} \\ 
        \beta(p^2-\mu/2)  & {\rm for\ bosons}
      \end{array}\right. 
\end{equation}
for $0\leq s\leq t$. For $-1\leq s\leq 0$ we use the lower bounds in
(\ref{q3cz}) and (\ref{q3czb}), respectively. It is then easy to see
that in this case 
\begin{equation}\label{lb3}
(1+s)h^{\rm
      F,B}_0(p)-s h^{\rm F,B}_q(p) \geq \beta \left[
      \min\{p^2,(p-q)^2,(p+q)^2\} -\mu\right]\,.
\end{equation}
Applying the bounds (\ref{lb3}) and (\ref{lb2}) to the denominator in
(\ref{sum2f}) and using (\ref{q2cz}) and (\ref{q2czb}), respectively,
to bound the expression $(h_0^{\rm F,B}(p) - h_q^{\rm F,B}(p))^2$ from above, we obtain that 
\begin{equation}\label{lbf}
(\ref{sumpq4}) \leq C_z |\Lambda| \beta^{1/2} q^4
\end{equation}
as long as $\beta q^2 \leq \const$ Here we have also used that $t\leq
1$ by definition. (Again, as in Section~\ref{lowsect}, we abuse the notation slightly and denote
by $C_z$ any expression that depends only on $z$.)

It remains to show that (\ref{lbf}) holds also for large values of
$\beta q^2$. To do this, we can go back to (\ref{sumpq4}) and
apply the bounds above directly to this term. In case $h_q^{\rm
  F,B}(p)\geq h_0^{\rm F,B}(p)$, we use (\ref{lb2}) (with $s=t$) as
well as the upper bounds in (\ref{q2cz}) and (\ref{q2czb}).
For the case $h_q^{\rm
  F,B}(p)\leq h_0^{\rm F,B}(p)$, we use (\ref{lb3}) and the lower
bounds in (\ref{q3cz}) and (\ref{q3czb}). We then split the sum into three
regions according to where the minimum in (\ref{lb3}) is attained, and
change variables from $p$ to $p-q$ or $p+q$, respectively. In this way we see that 
\begin{equation}
(\ref{sumpq4}) \leq C_z |\Lambda| \beta^{-1}| q|\left( 1 + \beta^{1/2}
  |q|\right) 
\end{equation}
for any value of $q$. Hence, in particular, (\ref{lbf}) holds for all
$q$. 

We have thus shown that
\begin{equation}
S(\Gamma,\Gamma_{0,q})\leq 2\left( 1 + C'_z \beta q^2\right) S(\Gamma,\Gamma_0) + C_z |\Lambda|
\beta^{1/2} q^4\,,
\end{equation}
with $C_z' = 1+2D_z$ for fermions and $C_z'=-1/\ln z$ for bosons.
We insert this bound into (\ref{qsum}) and sum over $q$. We can use
\begin{equation}\label{4der}
\frac 1{|\Lambda|}\sum_{q\in \frac{2\pi}L \Z^3} \widehat \eta_d(q)q^4 = \Delta^2\eta(0)
d^{-4} 
\end{equation}
and similarly for $q^4$ replaced by $q^2$. 
This leads to the result that, irrespective of whether we consider Fermi or Bose symmetry,
\begin{equation}\label{pps2}
S(\Gamma,\Gamma_d)\leq  C_z \left[ \left(1+\beta d^{-2}\right) S(\Gamma,\Gamma_0) + |\Lambda| \frac{\beta^{1/2}}{d^4}\right] \,,
\end{equation}
with $C_z$ a constant depending only on $z=e^{\beta \mu}$.

\subsection{Final Steps in the Proof}
If $n_{r,\xi}$ denotes the operator that counts the
number of particles in a ball of radius $r$ centered at $\xi$, we want a
lower bound on the expression
\begin{equation}
\int_{\R^3} d\xi\, \Tr\left[
n_{r,\xi}\left(n_{r,\xi}-1\right) \Gamma_{\chi_{r,\xi}}\right]\,. \label{srp}
\end{equation}
For a lower bound, we can replace the positive operator $n_{r,\xi}(n_{r,\xi}-1)$ by
$f_K(n_{r,\xi}(n_{r,\xi}-1))$, where 
\begin{equation}
f_K (t) = \left\{ \begin{array}{ll} t & {\rm for\ } t\leq K\\ K & {\rm
      for\ } t > K  
\end{array}\right.
\end{equation}
for some $K>0$ to be determined.
Then
\begin{align}\nonumber
\Tr\left[
n_{r,\xi}\left(n_{r,\xi}-1\right) \Gamma_{\chi_{r,\xi}}\right] &\geq \Tr\left[
f_K\left(n_{r,\xi}(n_{r,\xi}-1)\right) \Gamma_{\chi_{r,\xi}}\right]\\ \nonumber &\geq  \Tr\left[
f_K\left(n_{r,\xi}(n_{r,\xi}-1)\right) \Gamma_{d,\chi_{r,\xi}}\right] \\ & \quad - K
\|\Gamma_{\chi_{r,\xi}} -  \Gamma_{d,\chi_{r,\xi}}\|_1\,. \label{lst}
\end{align}
Next we note that $t-f_K(t)=[t-K]_+ \leq t^2/(4K)$, and hence
\begin{align}\nonumber
&\Tr\left[
f_K\left(n_{r,\xi}(n_{r,\xi}-1)\right) \Gamma_{d,\chi_{r,\xi}}\right] \\ &
\geq \Tr\left[
n_{r,\xi}(n_{r,\xi}-1) \Gamma_{d,\chi_{r,\xi}}\right] - \frac 1{4K} \Tr\left[
n_{r,\xi}^2(n_{r,\xi}-1)^2 \Gamma_{d,\chi_{r,\xi}}\right]\,. \label{qflt}
\end{align}
Note that $\Gamma_{d,\chi_{r,\xi}}$ is a
quasi-free state. Hence (compare with (\ref{fixrxi})--(\ref{intxi}))
\begin{align}\nonumber
&\int_{\R^3} d\xi\, \Tr\left[
n_{r,\xi}(n_{r,\xi}-1) \Gamma_{d,\chi_{r,\xi}}\right] \\& =
\int_{\Lambda\times\Lambda} dx\, dy\,
J_r(x-y)\left[\bar\rho^2\mp\mbox{$\sum_\sigma$}|\gamma_d(x,\sigma;y,\sigma)|^2\right]\,.
\end{align}
Moreover, the last term in (\ref{qflt}) is easy to estimate. Since
$\Gamma_{d,\chi_{r,\xi}}$ is quasi-free, it can be 
the explicitly expressed in terms of $\chi_{r,\xi}\gamma_d\chi_{r,\xi}$. A
simple estimate then yields, in the fermionic case,   
\begin{equation}
\Tr\left[
n_{r,\xi}^2(n_{r,\xi}-1)^2 \Gamma_{d,\chi_{r,\xi}}\right] \leq 
\left(\tr\left[\chi_{r,\xi}\gamma_d\right]\right)^2
  \left(\tr\left[\chi_{r,\xi}\gamma_d\right]+2\right)^2\,.
\end{equation}
In the bosonic case, we obtain
\begin{equation}
\Tr\left[
n_{r,\xi}^2(n_{r,\xi}-1)^2 \Gamma_{d,\chi_{r,\xi}}\right] \leq 24 
\left(\tr\left[\chi_{r,\xi}\gamma_d\right]\right)^2
  \left(\tr\left[\chi_{r,\xi}\gamma_d\right]+\half\right)^2\,.
\end{equation}
Note that $\tr[\chi_{r,\xi}\gamma_d]=4\pi r^3 \bar\rho /3$ as long as
$\Lambda$ contains the ball of radius $r$ centered at $\xi$, since $\gamma_d$ has a
constant density $\bar\rho$. For any $\xi$ and $r$ we have
$\tr[\chi_{r,\xi}\gamma_d]\leq 4\pi r^3 \bar\rho /3$. Integrating over $\xi$
thus yields
\begin{equation}
\int_{\R^3} d\xi\, \Tr\left[
n_{r,\xi}^2(n_{r,\xi}-1)^2 \Gamma_{d,\chi_{r,\xi}}\right] \leq \const |\Lambda|
\left( r^3\bar\rho\right)^2 \left( 1 + r^3\bar\rho\right)^2\,.
\end{equation} 

To estimate the last term in (\ref{lst}), we first note that
$\|\Gamma_{\chi_{r,\xi}} - \Gamma_{d,\chi_{r,\xi}}\|_1^2 \leq 2
S(\Gamma_{\chi_{r,\xi}},\Gamma_{d,\chi_{r,\xi}})$
\cite[Thm.~1.15]{ohya}.
Using Schwarz's inequality for the
$\xi$-integration yields
\begin{equation}\label{ad1}
\int_{\R^3} d\xi\, \|\Gamma_{\chi_{r,\xi}} - \Gamma_{d,\chi_{r,\xi}}\|_1\leq
\sqrt 2 (L+2r)^{3/2} \left( \int_{\R^3} d\xi\,
  S(\Gamma_{\chi_{r,\xi}},\Gamma_{d,\chi_{r,\xi}}) \right)^{1/2}\,.
\end{equation}
Here we have also used the fact that the integrand is zero if the distance between $\xi$ and $\Lambda$ is bigger than
$r$,  since there are no particles outside
$\Lambda$ and hence both restricted states are the
Fock space vacuum in this case. To estimate the last term in
(\ref{ad1}), we would like to use  (\ref{hest}). We note that, again
by monotonicity of the relative entropy,
$S(\Gamma_{\chi_{r,\xi}},\Gamma_{d,\chi_{r,\xi}})\leq
S(\Gamma_{\chi^\per_{r,\xi}},\Gamma_{d,\chi^\per_{r,\xi}})$. The
latter quantity is periodic in $\xi$, with period $L$. Moreover,
since $r\leq L/2$ by assumption, the cube of side length $L+2r$ is
contained within $3^3$ copies of $\Lambda$, and hence
\begin{equation}
\int_{\R^3} d\xi\,
  S(\Gamma_{\chi_{r,\xi}},\Gamma_{d,\chi_{r,\xi}}) \leq 3^3
  \int_{\Lambda} d\xi\,
  S(\Gamma_{\chi^\per_{r,\xi}},\Gamma_{d,\chi^\per_{r,\xi}}) \,. 
\end{equation}
Using (\ref{hest}) this yields
\begin{equation}
\int_{\R^3} d\xi\, \|\Gamma_{\chi_{r,\xi}} - \Gamma_{d,\chi_{r,\xi}}\|_1\leq
4 (L+2r)^{3/2} (3\bar d)^{3/2} S(\Gamma,\Gamma_d)^{1/2}\,.
\end{equation}
Note that $(L+2r)\leq (3/2)|\Lambda|^{1/3}$ since $2r\leq L/2$ by
assumption, as well as $\bar d \leq 2 d$.

Collecting all the terms and optimizing over $K$, we obtain the lower bound
\begin{align}\nonumber
&\int_{\R^3} d\xi\, \Tr\left[ n_{r,\xi}\left(n_{r,\xi}-1\right)
\Gamma\right] \\ \nonumber &\geq \int_{\Lambda\times\Lambda} dx\, dy\, J_r
(x-y) \left[ \bar\rho^2 \mp\mbox{$\sum_\sigma$} |\gamma_d(x,\sigma;y,\sigma)|^2\right] \\ & \quad
- \const r^3\bar\rho \left(1+ r^3\bar\rho\right)|\Lambda|^{3/4} d^{3/4}
S(\Gamma,\Gamma_d)^{1/4}\,. \label{final}
\end{align}
Note that $|\gamma_d(x,\sigma;y,\sigma)|\leq |\gamma_0(x,\sigma;y,\sigma)|$ because of
(\ref{defgd}) and the fact that $|\eta^\per_d|\leq 1$. Hence (\ref{final}), together with
(\ref{pps2}), proves the Theorem in the fermionic case.

In the bosonic case, we have to estimate, in addition, the term
\begin{equation}\label{bosadd}
\mbox{$\sum_\sigma$} \int_{\Lambda\times\Lambda} dx\, dy\, J_r
(x-y)   |\gamma_0(x,\sigma;y,\sigma)|^2(1-\eta^\per_d(x-y)^2)\,.
\end{equation}
We use that $J_r(x)\leq (4\pi/3)r^3$ and
$|\gamma_0(x,\sigma;y,\sigma)|\leq \bar\rho /n $. Moreover, we can
estimate 
$\eta^\per_d(x)^2\geq 1-\const (x/d)^\vu$ for any $0<\vu\leq 2$. Choosing
$\vu=1/4$ we obtain the bound
\begin{equation}
(\ref{bosadd}) \leq \const r^3\bar\rho\, d^{-1/4}
\int_{\Lambda\times\Lambda} dx\, dy\,  |\gamma_0(x,\sigma;y,\sigma)| |x-y|^{1/4}\,.
\end{equation}
By simple scaling, the integral is bounded above by
$C_z|\Lambda|\beta^{1/8}$ for some $z$-dependent constant. Hence the
error term (\ref{bosadd}) can be absorbed into the error terms already present in (\ref{final}), merely adjusting the constant. 

This finishes the proof of Theorem~\ref{P1}.

\appendix

\section{Appendix}
\begin{proof}[Proof of Lemmas~\ref{L1} and~\ref{L2}]
  We first prove (\ref{q3cz}) and (\ref{q3czb}). Since both $x\mapsto
  \ln[(2-x)/x]$ and $x\mapsto \ln[(2+x)/x]$ are monotone decreasing
  (for $0<x<2$ and $x>0$, respectively), we can obtain upper and lower
  bounds on $h^{\rm F,B}_q(p)$ by
  replacing $\gamma^{\rm F,B}_0(p+q)$ and $\gamma^{\rm F,B}_0(p-q)$ by the minimal and
  maximal value of these two expressions, respectively. This yields
  (\ref{q3cz}) and (\ref{q3czb}). 
  
  The upper bound in (\ref{q2czb}) follows immediately from convexity
  of the map $x\mapsto \ln[(2+x)/x]$ for $x>0$.
  
  The proof of (\ref{q2cz}) and the lower bound in (\ref{q2czb}) is a
  bit more tedious, but elementary. For convenience we set $\beta=1$,
  the correct $\beta$-dependence follows easily by scaling. For $0\leq
  \lambda\leq 1$, we define $f(\lambda)=h^{\rm F,B}_{\lambda q}(p)$.
  Note that $f'(0)=0$ and hence
\begin{equation}\label{useb}
h^{\rm F,B}_q(p)-h^{\rm F,B}_0(p) = f(1)-f(0)= \int_0^1 d\lambda\, (1-\lambda) f''(\lambda)\,.
\end{equation}
To calculate $f''(\lambda)$ it is useful to note that 
\begin{equation}
q\nabla \gamma^{\rm F,B}_0(p)=-2 p q\, 
\gamma^{\rm F,B}_0(p)\big(1\mp\gamma^{\rm F,B}_0(p)\big) 
\end{equation}
and 
\begin{align}
(q\nabla)^2 \gamma^{\rm F,B}_0(p) = & -2 q^2 \gamma^{\rm
  F,B}_0(p)\big(1\mp\gamma^{\rm F,B}_0(p)\big) \\ \nonumber & + 8
(p q)^2 \gamma^{\rm F,B}_0(p)\big(1\mp\gamma^{\rm F,B}_0(p)\big)\big(\half \mp \gamma^{\rm F,B}_0(p)\big)\,.
\end{align}
Denoting $p_{\pm}=p\pm \lambda q$ and $\gamma_\pm=\gamma^{\rm F,B}_0(p_\pm)$, we
therefore have
\begin{align}
  f''(\lambda)&=\left[-\frac{1}{\left(2\mp\gamma_+\mp\gamma_-\right)^2}+
    \frac 1{(\gamma_++\gamma_-)^2}\right] \\ \nonumber &\qquad \times
  \Big(2 p_+ q\,
  \gamma_+(1\mp\gamma_+) -2 p_- q\, \gamma_-(1\mp\gamma_-)\Big)^2 \\
  \nonumber &- \left[\pm\frac{1}{2\mp\gamma_+\mp\gamma_-}+ \frac
    1{\gamma_++\gamma_-}\right] \Big(-2q^2 \gamma_+(1\mp\gamma_+)-2 q^2
  \gamma_-(1\mp\gamma_-) \\ \nonumber &\qquad + 8(p_+ q)^2
  \gamma_+(1\mp\gamma_+)(\half\mp\gamma_+) + 8(p_- q)^2
  \gamma_-(1\mp\gamma_-)(\half\mp\gamma_-)\Big)\,.
\end{align}
Rearranging the various terms we can write
\begin{align}\label{secli}
f''(\lambda)&= \frac {\pm 4}{\gamma_++\gamma_-}\Biggl[ (p_+q)^2
  \gamma_+^2(1\mp\gamma_+) + (p_-q)^2 \gamma_-^2(1\mp\gamma_-) \\
  \nonumber & \qquad\qquad\qquad\qquad \mp
  \frac{\gamma_+\gamma_-}{\gamma_++\gamma_-}\Big(p_+q (1\mp\gamma_+)+
  p_-q(1\mp\gamma_-)\Big)^2 \Biggl]
\\ \nonumber &\mp \frac 4{2\mp \gamma_+\mp\gamma_-}\Biggl[(p_+q)^2
\gamma_+(1\mp\gamma_+)^2+ (p_-q)^2
\gamma_-(1\mp\gamma_-)^2 \\ \nonumber & \qquad\qquad\qquad\qquad \mp
\frac{(1\mp\gamma_+)(1\mp\gamma_-)}{2\mp\gamma_+\mp\gamma_-}\Big(p_+q
\,\gamma_++  p_-q\,\gamma_-\Big)^2\Biggl] 
\\ \nonumber &+  \left[\pm\frac{1}{2\mp\gamma_+\mp\gamma_-}+ \frac
    1{\gamma_++\gamma_-}\right] \Big(2q^2 \gamma_+(1\mp\gamma_+)+2 q^2
  \gamma_-(1\mp\gamma_-)\Big)\,. 
\end{align}
The term in the last line is positive and bounded above by $4q^2$,
both in the fermionic and bosonic case. 
For an upper bound in the fermionic case, we use that $p_\pm^2\gamma_\pm\leq D_z$,
$|p_\pm|\gamma_\pm\leq \sqrt{D_z}$ as well as $0\leq \gamma_\pm\leq 1$
to get $f''(\lambda)\leq 4q^2(1+2D_z)$. Similarly we can obtain a
lower bound. Using that $p_+q + p_-q = 2 pq$ in the second line in
(\ref{secli}), a simple estimate yields $f''(\lambda)\geq -12 q^2 D_z
- 8 p^2 q^2$ in the fermionic case. Using these bounds in (\ref{useb})
proves (\ref{q2cz}).

In the bosonic case, we only need to prove a lower bound on
(\ref{secli}). Proceeding as above, this time using $p_\pm^2
\gamma_\pm(1+\gamma_\pm)\leq D_z$ and $|p_\pm|\gamma_\pm \leq
\sqrt{D_z}$ we obtain again $f''(\lambda)\geq -12 q^2 D_z
- 8 p^2 q^2$. This finishes the proof of the lemmas.
\end{proof}

\end{document}